\documentclass[aps,pra, twocolumn]{revtex4}
\usepackage{graphicx}    
\usepackage{amsmath}

\begin{document}

\title{Temporal second-order coherence function for displaced-squeezed thermal states}

\author{Moorad Alexanian}
\email[]{alexanian@uncw.edu}

\affiliation{Department of Physics and Physical Oceanography\\
University of North Carolina Wilmington\\ Wilmington, NC
28403-5606\\}

\date{\today}

\begin{abstract}
We calculate the quantum mechanical, temporal second-order coherence function for a single-mode, degenerate parametric amplifier for a system in the Gaussian state, viz., a displaced-squeezed thermal state. The calculation involves first the dynamical generation at time $t$ of the Gaussian state from an initial thermal state and subsequent measurements of two photons a time $\tau \geq 0$ apart. The generation of the Gaussian state by the parametric amplifier ensures that the temporal second-order coherence function depends only on $\tau$, via $\tau/t$, for given Gaussian state parameters, Gaussian state preparation time $t$, and average number $\bar{n}$ of thermal photons. It is interesting that the time evolution for displaced thermal states shows a power decay in $\tau/t$ rather than an exponential one as is the case for general, displaced-squeezed thermal states.\\

\end{abstract}

\maketitle {}

\section{Introduction}

The field of quantum computation and quantum information, as applied to quantum computers, quantum cryptography, and quantum teleportation, was originally based on the manipulation of quantum information in the form of discrete quantities like qubits, qutrits, and higher-dimensional qudits. Nowadays the emphasis has shifted on processing quantum information by the use of continuous-variable quantum information carriers. In this regard, use is now made of any combination of Gaussian states, Gaussian operations, and Gaussian measurements \cite{WPP12, ARL14}. The interest in Gaussian states is both theoretical as well as experimental since simple analytical tools are available and, on the experimental side, optical components effecting Gaussian processes are readily available in the laboratory \cite{WPP12}.

The question of how a quantum computer can be used to simulate experiments on quantum systems in thermal equilibrium has been addressed by preparing states that, after long enough time, approach the desired equilibrium. The latter allows estimating efficiently equilibrium time-correlation functions on a quantum computer \cite{TD00}.

Methods have been proposed for quantum algorithms for efficiently computing $n$-time correlation functions for bosonic systems. Note that all quantum mechanical time evolution is governed by the Hamiltonian and, in most cases, the extraction of the time evolution of the system and the determination of its stationary states, albeit theoretically straightforward, may not be easily accomplished \cite{PDECS14}.

Quantum optical systems give rise to interesting nonclassical behavior such as photon antibunching and sub-Poissonian photon statistics owing to the discreetness or photon nature of the radiation field \cite{SZ97}. These nonclassical features can best be quantified with the aid of the temporal second-order quantum mechanical correlation function $g^{(2)}(\tau)$ and experimentally studied using a Hanbury Brown--Twiss intensity interferometer modified for homodyne detection \cite{GSSRL07}.

Physical realizations and measurements of the second-order coherence function $g^{(2)}(\tau)$ of light have been studied earlier via a degenerate parametric amplifier (DPA) \cite{KHM93,LO02,GSSRL07}. Measurements of the normalized, second-order intensity correlation function for a stationary process of light emitted from a cavity QED system composed of $N$, two-level atoms coupled to a single mode of the electromagnetic field has been shown to exhibit the nonclassical features and dynamics of the atom-field interaction \cite{FMO00}.

This paper is organized as follows. In Sec. II, we consider the general Hamiltonian of the DPA and its relation and equivalence to the unitary transformation of creation and annihilation operators that give rise to displaced and squeezed states. Section III considers the second-order coherence function for Gaussian states, displaced-squeezed thermal states, with special attention of how the DPA Hamiltonian is used to both dynamically generate a Gaussian state from an initial thermal state and the temporal behavior of the two-time coherence function without assuming the field to be statistically stationary. In Sec. IV, we consider particular cases of the time evolution of the coherence function for displaced thermal states and for squeezed thermal states. It is interesting that the former show a power-law decay in time while the latter decay exponentially. Finally, Sec. V summarizes and compares our results with known calculations.

\section{General Gaussian states and Degenerate Parametric Amplification}

Consider the unitary transformations
\begin{equation}
\hat{S}(-\xi)\hat{D}(-\alpha) \hat{a} \hat{D}(\alpha)\hat{S}(\xi) = \hat{a} \cosh r - \hat{a}^{\dag} \exp(i\theta) \sinh r +\alpha,
\end{equation}
and
\begin{equation}
\hat{S}(-\xi)\hat{D}(-\alpha) \hat{a}^{\dag} \hat{D}(\alpha)\hat{S}(\xi) = \hat{a}^{\dag} \cosh r - \hat{a} \exp(-i\theta) \sinh r +\alpha^*,
\end{equation}
with  the displacement $\hat{D}(\alpha)= \exp{(\alpha \hat{a}^{\dag} -\alpha^* \hat{a})}$  and the squeezing $\hat{S}(\xi)=  \exp\big{(}-\frac{\xi}{2} \hat{a}^{\dag 2} + \frac{\xi^*}{2} \hat{a}^{2} \big{ )}$ operators, where $\hat{a}$ ($\hat{a}^{\dag})$ is the photon annihilation (creation) operator, $\xi = r \exp{(i\theta)}$, and $\alpha= |\alpha|\exp{(i\varphi)}$.

The Hamiltonian for degenerate parametric amplification, in the interaction picture, is
\begin{equation}
\hat{H} = c \hat{a}^{\dag 2} + c^* \hat{a}^2 + b\hat{a} + b^* \hat{a}^\dag.
\end{equation}
One obtains, with the aid of the Baker-Campbell-Hausdorff formula, viz., $\exp{[i\hat{H}\tau/\hbar]}\hspace{0.03in} \hat{a}\hspace{0.02in} \exp{[-i\hat{H}\tau/\hbar]} =
\hat{a} + [i\hat{H}\tau/\hbar,\hat{a}] + (1/2!) [i\hat{H}\tau/\hbar, [i\hat{H}\tau/\hbar,\hat{a}]] +\cdots$, that
\[
\exp{[i\hat{H}\tau/\hbar]} \hat{a} \exp{[-i\hat{H}\tau/\hbar]} =  \hat{a}\cosh(\Omega\tau)
\]
\begin{equation}
-i\hat{a}^\dag \exp{(i\chi)} \sinh(\Omega\tau) -ib^* \frac{\sinh (\Omega\tau)}{2|c|}
\end{equation}
\[
+ b\exp{(i\chi)} \frac{\cosh (\Omega\tau)-1 }{2|c|},
\]
where
\begin{equation}
c= |c|\exp{(i\chi)} \hspace{0.3in}  \textup{and }\hspace{0.3in} \Omega = 2|c|/\hbar.
\end{equation}

We define $\xi(\tau)$ and $\alpha(\tau)$ by
\begin{equation}
\xi(\tau) = \frac{2i c}{\hbar}\tau
\end{equation}
and
\begin{equation}
\alpha (\tau)= -ib^* \frac{\sinh (\Omega\tau)}{2|c|} + b\exp{(i\chi)} \frac{\cosh (\Omega\tau) -1 }{2|c|}.
\end{equation}
Note that $\xi(0)=\alpha(0)=0$.

We choose the displacement parameters of the Gaussian state $\alpha = \alpha(t)$ and the squeezing parameter $\xi=\xi(t)$. Eqs. (6) and (7) for $\xi(t)$ and $\alpha(t)$ can be inverted and thus give the values of $tc$ and $tb$ in terms of the parameters $\alpha$ and $\xi$ of the Gaussian state (see Eqs. (20) and (21) below).  We then have that $\exp{[i \hat{H}t/\hbar]} \hat{a} \exp{[-i \hat{H} t/\hbar]}= \hat{a}\cosh r -\hat{a}^\dag\exp{(i\theta)}\sinh r+\alpha$, which is equal to  the right-hand-side of Eq. (1), where $\hat{H}$ is the Hamiltonian (3) with $tc$ and $tb$ given in terms of $\alpha$ and $\xi$ (see Eqs. (20) and (21) below).

\section{Temporal second-order correlation function}

The quantum mechanical temporal second-order coherence \cite{SZ97} is given by
\begin{equation}
g^{(2)}(\tau) =\frac{\langle \hat{a}^{\dag}(t) \hat{a}^{\dag}(t+\tau) \hat{a}(t+\tau) \hat{a}(t)\rangle }{\langle \hat{a}^{\dag}(t)  \hat{a}(t)\rangle \langle \hat{a}^{\dag}(t+\tau)\hat{a}(t+\tau) \rangle},
\end{equation}
with $\tau\geq 0$, where the operators are in the interaction picture and the expectation values, representing an average over the initial photon states, are determined by the thermal density matrix
\begin{equation}
\hat{\rho_{0}} = \exp{(-\beta \hbar \omega\hat{n})}/ \textup{Tr}[\exp{(-\beta \hbar \omega \hat{n})}],
\end{equation}
where $\hat{n}= \hat{a}^{\dag} \hat{a}$, $\beta=1/(k_{B}T)$, and $\bar{n}= \textup{Tr}[\hat{\rho}_{0} \hat{n}]$. In quantum optics, one usually deals with statistically stationary fields, viz., the correlation functions of the field are invariant under time translation and so the correlation functions, for instance, the temporal second-order coherence (8), would be a function of the single variable $\tau$. Note, however, this requires that the thermal density matrix (9) commutes with the Hamiltonian (3) that governs the time development of the system, which is not generally so.

It is important to remark that Eq. (8) is the simpler form of the quantum-mechanical second-order degree of coherence when the radiation field consists of only a single mode. This form follows directly from the general definition of the quantum-mechanical second-order degree of coherence given by Eq. (4.2.19) in \cite {SZ97}, which is the same as that given by Eq. (9.38) in \cite{RL73}. Note that our Eq. (8) differs from that given by Eq. (4.2.21) in \cite {SZ97} in the denominator, where in Eq. (8) we have not assumed the field to be statistically stationary.

We do not suppose statistically stationary fields, instead, in our scheme, the system is initially in the thermal state $\hat{\rho}_{0}$. After time $t$, the system evolves to the Gaussian state $\hat{\rho}_{G}$ and a photon is annihilated at that time $t$, the system then  develops in time and after a time $\tau$ another photon is annihilated. Therefore, two photon are annihilated in a time separation $\tau$ when the system is in the Gaussian density state $\hat{\rho}_{G}$. The second-order coherence function is given by
\begin{equation}
g^{(2)}(\tau) = \frac{\langle \hat{a}^{\dag}(0) \hat{a}^{\dag}(\tau) \hat{a}(\tau) \hat{a}(0)\rangle }{\langle \hat{a}^{\dag}(0)  \hat{a}(0)\rangle \langle \hat{a}^{\dag}(\tau)\hat{a}(\tau) \rangle},
\end{equation}
with the aid of the trace property $Tr(\hat{A}\hat{B}\hat{C})=Tr(\hat{B}\hat{C}\hat{A})$ and since $\hat{a}(t)=\exp(i\hat{H}t/\hbar)\hat{a}(0)\exp(-i\hat{H}t/\hbar)= \hat{S}(-\xi)\hat{D}(-\alpha) \hat{a}(0) \hat{D}(\alpha)\hat{S}(\xi)$, where $\xi = \xi(t)$, $\alpha=\alpha(t)$ and  $e^{i\theta} = i e^{i\chi}$, which follows from Eqs. (4) and (7).  All the expectation values in (10) and henceforth are traces with the Gaussian density operator, viz., a displaced-squeezed thermal state, given by
\begin{equation}
\hat{\rho}_{G}=  \hat{D}(\alpha) \hat{S}(\xi)\hat{\rho}_{0} \hat{S}(-\xi) \hat{D}(-\alpha),
\end{equation}
where $r= \Omega t$, $\xi = \xi(t)$ by Eq. (6), and $\alpha=\alpha(t)$ by Eqs. (4) and (7). Expression (10) for the second-order coherence function follows from Eq. (8) since $\hat{a}(t+\tau ) =\hat{S}(-\xi)\hat{D}(-\alpha) \hat{a}(\tau) \hat{D}(\alpha)\hat{S}(\xi)$ for $\tau\geq 0$. Therefore, $\textup {Tr}[\hat{\rho}_{0}\hat{a}^\dag(t) \hat{a}(t)]/\textup{Tr} [ \hat{\rho}_{0}]=  \textup {Tr}[\hat{\rho}_{G}\hat{a}^\dag(0) \hat{a}(0)]/\textup{Tr} [ \hat{\rho}_{G}]$,

 $\textup {Tr}[\hat{\rho}_{0}\hat{a}^\dag(t+\tau) \hat{a}(t+\tau)]/\textup{Tr} [ \hat{\rho}_{0}]=  \textup {Tr}[\hat{\rho}_{G}\hat{a}^\dag(\tau) \hat{a}(\tau)]/\textup{Tr} [ \hat{\rho}_{G}]$, $\textup {Tr}[\hat{\rho}_{0}\hat{a}^\dag(t) \hat{a}^\dag(t+\tau)\hat{a}(t+\tau) \hat{a}(t)]/\textup{Tr} [ \hat{\rho}_{0}]=  \textup {Tr}[\hat{\rho}_{G}\hat{a}^\dag(0)\hat{a}^\dag(\tau) \hat{a}(\tau) \hat{a}(0)]/\textup{Tr} [ \hat{\rho}_{G}]$, and $ \textup {Tr}[\hat{\rho}_{G}] = \textup {Tr}[\hat{\rho}_{0}]$. The coherence function $g^{(2)}(\tau)$ is a function of $\Omega \tau=(\tau/t) r$, $\alpha$, $\xi$, and $\bar{n}$, where the preparation time $t$ is the time that it takes the system to dynamically generate the Gaussian density $\hat{\rho}_{G}$ given by (11) from the initial thermal state $\hat{\rho}_{0}$ given by (9).

Now
\begin{equation}
\langle \hat{a}^{\dag}(\tau)\hat{a}(\tau) \rangle = (\bar{n}+1/2)\cosh[2(r+ \Omega \tau)]-1/2 + |A(\tau)|^2,
\end{equation}
where
\begin{equation}
A(\tau)= \alpha\cosh (\Omega \tau) -i \alpha^* \exp{(i\theta)} \sinh (\Omega \tau) +\alpha(\tau),
\end{equation}
with $\Omega$ and $\alpha(\tau)$ defined by Eqs. (5) and (7), respectively. One has for $\tau=0$ that
\begin{equation}
\langle \hat{a}^{\dag}(0)\hat{a}(0)\rangle =(\bar{n}+1/2 )\cosh(2r) -1/2 + |\alpha|^2.
\end{equation}
The expression for $|A(\tau)|^2$  is given by (A.2) in the Appendix.

A lengthy albeit straightforward calculation of the temporal second-order correlation function (10) gives
\begin{equation}
g^{(2)}(\tau)= 1 + \frac{n^2(\tau)+ s^2(\tau) +u(\tau) n(\tau) -v(\tau)s(\tau)}{\langle \hat{a}^{\dag}(0)\hat{a}(0)\rangle \langle \hat{a}^{\dag}(\tau)\hat{a}(\tau) \rangle},
\end{equation}
where
\begin{equation}
n(\tau)= (\bar{n}+ 1/2)\cosh \big{(}2r+ \Omega\tau\big{)} -(1/2)\cosh (\Omega\tau),
\end{equation}

\begin{equation}
s(\tau)= (\bar{n}+ 1/2)\sinh \big{(}2r+\Omega\tau\big{)} -(1/2)\sinh (\Omega\tau),
\end{equation}

\begin{equation}
u(\tau) = \alpha A^*(\tau) + \alpha^* A(\tau),
\end{equation}
and

\begin{equation}
v(\tau)=\alpha A(\tau)\exp{(-i\theta)}+\alpha^* A^*(\tau)\exp{(i\theta)}.
\end{equation}
Note that number of photons (12) increases exponentially for $\Omega \tau \gg 1$ owing to the pumping of photons into the cavity; however, correlation (15) approaches a finite value for $\Omega \tau \gg 1$.

Result (15) reduces to the known results for $\tau=0$, for the vacuum state with $\bar{n}=0$  \cite {GSSRL07} and for the thermal state with $\bar{n}\neq 0$ \cite{LDC14}, since $A(0)=\alpha$ and provided $\theta=2 \varphi$, which to minimize intensity fluctuations it is always optimal to squeeze the amplitude quadrature. The quantity $u(\tau)n(\tau) -  v(\tau) s(\tau)$ is given by (A1) in Appendix A and so $g^{(2)}(\tau)$ is a function of $ \Omega \tau$, $\bar{n}$, $r$, $|\alpha|$, and $\theta-2\varphi$. Note that $\Omega = r/t$, where the preparation time $t$ is the time that it takes the system to dynamically generate the Gaussian density $\hat{\rho}_{G}$ given by (11) from the initial thermal state $\hat{\rho}_{0}$ given by (9).

Our result for $g^{(2)}(0)$, which follows from (15) with $\tau =0$, also agrees with Eq. (3.10) of the first of Ref. [6] for their case with $2 \arg(\alpha) -\arg(\zeta) = \pi$.

The correlation function (15) gives rise to $g^{(2)}(\tau)\rightarrow 3$ as $r\rightarrow \infty$ for $\tau\geq 0$. This asymptotic behavior has been observed experimentally \cite{GSSRL07} for $\tau =0$. It is interesting, however, that $g^{(2)}(\tau)\rightarrow 1$ as $|\alpha|\rightarrow \infty$ for $\tau\geq 0$, which is the result for a coherent state.

The parameters $c$ and $b$ in the degenerate parametric Hamiltonian (3) are determined by the parameters  $\alpha$ and $\xi$ of the Gaussian density of state (11) once the time $t$ is chosen that determines the time it takes for the system governed by the Hamiltonian (3) to generate the Gaussian density of state $\hat{\rho}_{G}$ from the initial density of state $\hat{\rho}_{0}$ and so
\begin{equation}
tc = -i\frac{\hbar}{2} r\exp(i\theta)
\end{equation}
and
\begin{equation}
tb= -i\frac{\hbar}{2}\Big{(} \alpha \exp{(-i\theta)} + \alpha^* \coth (r/2)\Big{)} r.
\end{equation}
Note that the frequency $\Omega = r/t$ is then determined as well.

\begin{figure}
\begin{center}
   \includegraphics[scale=0.3]{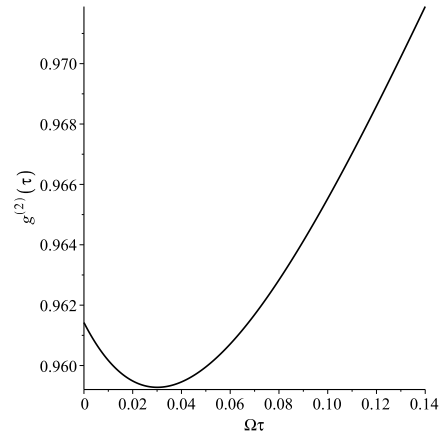}
\end{center}
\label{fig:theFig}
  \caption{Temporal second-order correlation function $g^{(2)}(\tau)$ for $\bar{n}=0.1$, $r=0.3$, and $|\alpha|=0.8$. The classical inequality (23) is violated for  $0<\Omega \tau < 0.0674$ and for $\Omega \tau > 0.593$ since $g^{(2)}(0.0674)= g^{(2)}(0)= 2-  g^{(2)}(0.593)$ and $g^{(2)}(\tau)$ is monotonically  increasing after achieving its minimum.  Note that the classical condition $g^{(2)}_{c}(0)\geq 1$ is violated since the field is sub-Poissonian. For $0\leq \Omega \tau \leq 0.0674$ one has photon bunching since $g^{(2)}(\tau)\leq g^{(2)}(0)$ and photon antibunching for $\Omega \tau> 0.0674$. The minimum occurs at $\Omega \tau =0.0300$ and $\lim_{\tau\rightarrow \infty } g^{(2)}(\tau)= 1.238$.}
\end{figure}

If the field operators in the second-order correlation function (8) are treated as c-numbers, then, with the aid of the Cauchy-Schwarz inequality, the classical correlation function $g^{(2)}_{c}(\tau)$ satisfies
\begin{equation}
g^{(2)}_{c}(0) \geq 1 \hspace{0.3in} \textup{and}    \hspace{0.3in} g^{(2)}_{c}(0) \geq  g^{(2)}_{c}(\tau).
\end{equation}
Similarly, one can derive the classical inequality \cite{RC88}
\begin{equation}
|g^{(2)}_{c}(0)-1| > |g^{(2)}_{c}(\tau)-1|,
\end{equation}
that is, $g^{(2)}_{c}(\tau)$ can never be farther away from unity than it was initially at $\tau=0$. In certain quantum optical systems, the inequality (22) may be violated with the result $g^{(2)}(\tau) > g^{(2)}(0)$. Another nonclassical inequality is given by  $g^{(2)}(0) <1$.

\begin{figure}
\begin{center}
   \includegraphics[scale=0.3]{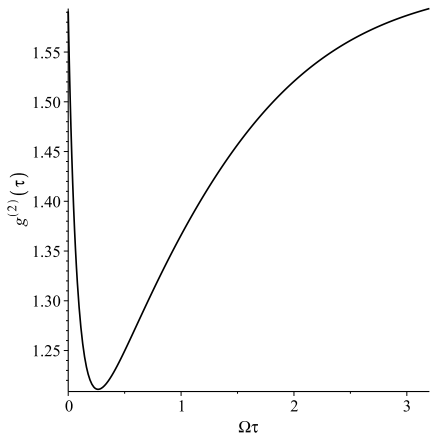}
\end{center}
\label{fig:theFig}
  \caption{Temporal second-order correlation function $g^{(2)}(\tau)$ for $\bar{n}=0$, $r=0.3$, and $|\alpha|=0.4$. The minimum occurs for $\Omega \tau = 0.264$ and $g^{(2)}(0.264)= 1.211$. The classical condition $g^{(2)}(0) \geq g^{(2)}(\tau)$ is valid for $0\leq \Omega \tau \leq 3.113$ and violated for $\Omega \tau > 3.113$. Similarly for the inequality (23). Note $g^{(2)}(0) = g^{(2)}(3.113)= 1.590$ and $\lim_{\tau\rightarrow \infty } g^{(2)}(\tau)= 1.624$. }
\end{figure}

Fig. 1 shows the mixed character of the correlation function $g^{(2)}(\tau)$ with both classical and quantum features.

Fig. 2 shows, as Fig. 1 does, mixed classical and quantum features with the quantum feature of photon antibunching occurring for $\Omega \tau > 3.113$ Note that $g^{(2)}(0) > 1$ and so we have a super-Poissonian field and in both fields depicted in Fig. 1 and Fig. 2, the quantum features occurs at later times except for Fig.1, where $g^{(2)}(0) <1$.

\section{particular correlation functions}

The general Hamiltonian of the degenerate parametric amplifier (3) gives rise to the time dependence (15) for $g^{(2)}(\tau)$. In this section, we consider two particular cases for the correlation function derived from (15). First, the case of the time evolution of the system for a strictly displaced thermal state, viz., $r=0$, and, secondly, for the time evolution of a strictly squeezed thermal state, viz., $\alpha =0$.

\subsection{Time evolution of displaced thermal state}

Consider the limit $r\rightarrow 0$  that implies, with the aid of (20) and (21), that the frequency $\Omega = r/t \rightarrow 0$ as $r\rightarrow 0$ and so the resulting correlation function is
\begin{equation}
g^{(2)}_{(r=0)}(\tau)= 1+ \frac{\bar{n}}{|\alpha|^2 +\bar{n}}\Bigg{(}\frac{2|\alpha|^2 (\tau/t +1) +\bar{n}}{|\alpha|^2(\tau/t+1)^2 +\bar{n}}\Bigg{)}.
\end{equation}
Result (24) satisfies both of the classical conditions (22) and (23) and so the correlation is strictly classical. The particle number is
\begin{equation}
\langle \hat{a}^{\dag}(\tau)\hat{a}(\tau) \rangle_{(r=0)} = |\alpha|^2 (\tau/t+1)^2 + \bar{n},
\end{equation}
which indicates that the number of particles (25) increases with time $\tau$. The correlation (24) approach unity as $\tau/t \rightarrow \infty$, which is the temporal correlation function for a coherent state, viz., a laser operating far above threshold. Also, for $\bar{n}\gg |\alpha|^2$ and $0\leq \tau/t<\infty$, the correlation (24) approaches the value of 2, which value corresponds to the thermal state.

It is interesting that for fixed $r>0$, the correlation (15) approaches its asymptotic value exponentially with $\Omega \tau$ for $\Omega \tau\gg 1$. However, for $r=0$, (24) gives rise to a sort of long-range order, which substantially optimizes correlations. This result is reminiscent of the seminal molecular dynamical simulations for hard spheres \cite{AW70} that showed that the two-time, velocity autocorrelation function decayed not exponentially in time but rather as a power of time as it approaches its asymptotic value.

\begin{figure}
\begin{center}
   \includegraphics[scale=0.3]{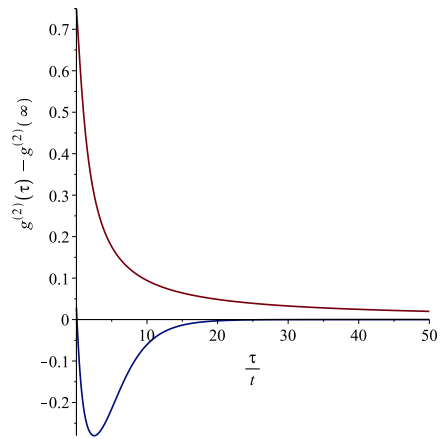}
\end{center}
\label{fig:theFig}
  \caption{Temporal second-order correlation function $g^{(2)}(\tau)$ as a function of $\tau /t$, where $t$ is the preparation time. Red graph: Eq. (24) with $\bar{n}=1$, $|\alpha|=1$, $g^{(2)}(0)=1.750$, and $g^{(2)}(\infty)=1$. Blue graph: Eq. (15) with  $\bar{n}=1$, $r=0.3$, $|\alpha|=1$, $g^{(2)}(0)=1.615$, and $g^{(2)}(\infty)=1.586$. Both correlation functions satisfy the classical correlation inequalities (22) and (23). Note the contrast between the power law decay (red) versus the exponential decay (blue).  }
\end{figure}

Fig. 3 and Fig. 4 show the slow decay of the correlation function $g^{(2)}_{(r=0)}(\tau)$ for given values of $\bar{n}$ and $|\alpha|$ as contrasted to the behavior of $g^{(2)}(\tau)$ for the same values of $\bar{n}$ and $|\alpha|$ but for $r>0$. The blue graph in Fig. 4, which has a minimum, has interesting mixed classical-nonclassical field features: For $0<\tau<\tau_{0}$, where $g^{(2)}(\tau_{0})=g^{(2)}(0)$, one has that  $1> g^{(2)}(0) >  g^{(2)}(\tau)$ and so inequality (23) is violated. For $\tau_{0}<\tau<\tau_{1}$, where $g^{(2)}(\tau_{1})=2-g^{(2)}(0)$, one has that $g^{(2)}(\tau) >  g^{(2)}(0)$ and inequality (23) is satisfied. Finally, for $\tau >\tau_{1}$, all three inequalities in (22) and (23) are violated.

\subsection{Time evolution of squeezed thermal state}

Consider next the limiting case when $\alpha =0$. One has from (15) that
\[
g^{(2)}_{(\alpha=0)}(\tau) = 1+ \Big{(}\big{[}(\bar{n}+1/2)\cosh(2r+\Omega \tau)-(1/2)\cosh(\Omega \tau)\big{]}^2
\]
\begin{equation}
+\big{[}(\bar{n}+1/2)\sinh(2r+\Omega \tau)-(1/2)\sinh(\Omega \tau)\big{]}^2\Big{)}
\end{equation}
\[
\Big{/}\Big{(}\big{[}(\bar{n}+1/2)\cosh(2r)-1/2\big{]}\big{[}(\bar{n}+1/2)\cosh(2r+2\Omega \tau)-1/2\big{]}\Big{)}.
\]
Result (26) satisfies $g^{(2)}_{(\alpha=0)}(\tau)\geq 1$ for $\tau\geq 0$ and so the classical condition $g^{(2)}_{c}(0)\geq 1$. However, (26) violates the second classical condition in (22), viz., $g^{(2)}_{c}(0) \geq  g^{(2)}_{c}(\tau)$, since for particular values of $\tau$ for some given squeezing parameter $r$ and
$\bar{n}>0$, $g^{(2)}_{(\alpha=0)}(\tau)> g^{(2)}_{(\alpha=0)}(0)$.

\begin{figure}
\begin{center}
   \includegraphics[scale=0.3]{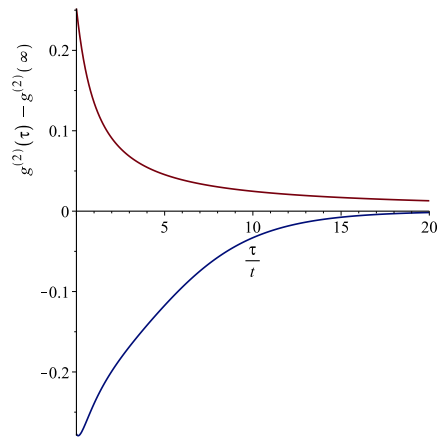}
\end{center}
\label{fig:theFig}
  \caption{Temporal second-order correlation function $g^{(2)}(\tau)$ as a function of $\tau /t$, where $t$ is the preparation time. Red graph: Eq. (24) with $\bar{n}=0.1$, $|\alpha|=0.8$, $g^{(2)}(0)=1.252$, and $g^{(2)}(\infty)=1$. Red correlation satisfies both classical inequalities (22) and (23). Blue graph: Eq. (15) with  $\bar{n}=0.1$, $r=0.3$, $|\alpha|=0.8$, $g^{(2)}(0)=0.961$ and $g^{(2)}(\infty)=1.238$. Minimum at $\tau=0.1002$. Define $g^{(2)}(0) = g^{(2)}(\tau_{0})=2-g^{(2)}(\tau_{1})$, where $\tau_{0}=0.2249$ and $\tau_{1}= 1.978$. Note the contrast between the power law decay (red) versus the exponential decay (blue).  }
\end{figure}

\begin{figure}
\begin{center}
   \includegraphics[scale=0.3]{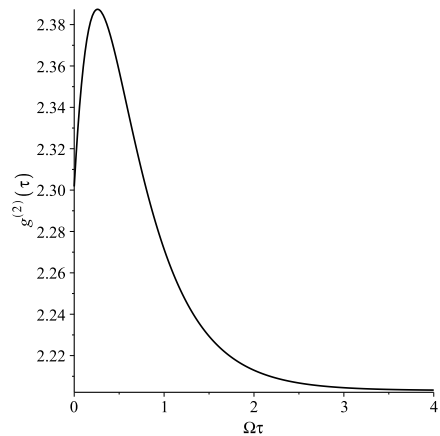}
\end{center}
\label{fig:theFig}
  \caption{Temporal second-order correlation function (26) for $g^{(2)}_{(\alpha=0)}(\tau)$ as a function of $\Omega \tau $ with $\bar{n}=1$ and $r=0.2$. One has     $g^{(2)}_{(\alpha=0)}(0)= g^{(2)}_{(\alpha=0)}(0.794)= 2.301$ and $g^{(2)}(\infty)=2.203$. Note that the first classical inequality (22) is satisfied, but the second inequality in (22) and inequality (23) are violated for $0 < \Omega \tau < 0.794$. Inequalities (22) and (23) are satisfied for $\tau \geq 0.794$ }
\end{figure}

Fig. 5 indicates the result for the preparation and subsequent time development of the correlation function for a squeezed thermal state. The field exhibit both classical and quantum features.

\section{summary and discussions}

We calculate the second-order coherence function (15) for Gaussian states, viz., displaced-squeezed thermal states, where the dynamics is governed solely by the general, degenerate parametric amplification Hamiltonian (3). Our result (15) for the correlation function is exact and is based on dynamically generating the Gaussian state first and subsequently determining the time evolution of the system without assuming the field to be statistically stationary.

For particular values of the parameters of the Gaussian state, we show plots that indicate the mixed classical/nonclassical nature of the correlation function. It is interesting that the behavior of the time evolution of the displaced thermal state, which is strictly classical, shows a decay in time not exponential, which is true for finite squeezed parameter, but rather as a power of time.

Some calculations of $g^{(2)}(\tau)$ include damping \cite{LDC14, MM93II} and so a direct comparison with our result (15) may be possible by setting the damping rate or damping constant equal to zero. It is important to remark that the Hamiltonian (3) for the general, degenerate parametric amplifier serves two purposes, viz., first to create the Gaussian state from an initial thermal state and secondly to give rise to the temporal, second-order coherence function.

The temporal evolution of our output correlations is due only to the intracavity dynamics governed by the Hamiltonian (3). In contrast, in calculating the time dependence of $g^{(2)}(\tau)$ in Ref. [6], use is made of commutation relations that depend on the filter shape used in the measurement process and so no direct comparison can be made with our result (15). It is interesting that the use of frequency filtering has been used to obtain frequency-resolved photon correlations that are used to optimize photon correlations \cite{GVL15}.

The weak interaction of a single-mode radiation field initially in a Gaussian state with a heat bath of arbitrary temperature preserves the Gaussian form of the characteristic function \cite{MM93II} and so the correlation function $g^{(2)}(\tau)$.  However, the time evolution is governed only by the damping and so setting the damping constant equal to zero yields only $g^{(2)}(0)$, which agrees with our result.

The calculation of the correlation function in Ref. (11) comes closest to ours since it adds to the general Hamiltonian of the degenerate  parametric amplifier a coupling to a single input/output waveguide characterized by a damping rate $\kappa$. The two-time intensity correlation $g^{(2)}(\tau)$ given by Eq. (10) of Ref. (11) for $\kappa=0$ does not reduce to our result (15). In fact, their $g^{(2)}(\tau)$ diverges exponentially for $\lambda \tau \gg 1$ and $\kappa\geq 0$, whereas our result (15) has a finite limit as $\Omega \tau\gg 1$.

\appendix
\section{Auxiliary functions}
Expression (15) for the correlation function $g^{(2)}(\tau)$ can be simplified with the aid of (7), (13), (20), and (21)
\[
u(\tau)n(\tau) -  v(\tau) s(\tau)
\]
\[
=|\alpha|^2 \Big{[}1+ \cosh (\Omega\tau) +\sinh (\Omega\tau)\coth (r/2)\Big{]}\cdot
\]
\begin{equation}
 \cdot\Big{[} n(\tau) - s(\tau) \cos (\theta-2\varphi)  \Big{]}
\end{equation}
\[
+|\alpha|^2 \Big{[}\sinh (\Omega\tau) - (\cosh (\Omega\tau)-1)\coth (r/2)\Big{]}\cdot
\]
\[
 \cdot \Big{[} n(\tau) \cos (\theta-2\varphi) - s(\tau)   \Big{]}
+2|\alpha|^2 n(\tau) \sinh (\Omega \tau) \sin(\theta-2\varphi).
\]
On can minimize the correlation function by choosing $\theta= 2\varphi$, which increases the negative over the positive terms in (A.1).

Now
\[
|A(\tau)|^2 =|\alpha|^2 \Bigg{|}\cosh(\Omega\tau)+\frac{1}{2}\coth(r/2) \sinh (\Omega \tau)
\]
\begin{equation}
-\frac{1}{2} (\cosh(\Omega \tau)-1)+\exp[i(\theta -2 \varphi)]\Big{[} -i \sinh(\Omega\tau)
\end{equation}
\[
+\frac{1}{2}\sinh(\Omega\tau)-\frac{1}{2}\coth(r/2)\big{(}\cosh(\Omega\tau)-1\big{)}\Big{]}\Bigg{|}^2.
\]

\section{Minimized correlation functions}

The correlation function (15) possesses a minimum value as a function of $|\alpha|$ for given $\Omega \tau$, $\bar{n}$, and $r$
\begin{equation}
|\alpha|^2 =-\frac{AD \pm \sqrt{D(AD-BE)(A-BC)}}{BD},
\end{equation}
where
\begin{equation}
A= n^2(\tau) +s^2(\tau),
\end{equation}
\[
B = \Big{(}1+\cosh(\Omega\tau)+ \coth(r/2)\sinh(\Omega\tau)+\sinh(\Omega\tau)
\]
\begin{equation}
-\coth(r/2)(\cosh(\Omega\tau)-1)\Big{)}\big{(}n(\tau)-s(\tau)\big{)},
\end{equation}
\begin{equation}
C = (\bar{n}+1/2)\cosh(2r)-1/2,
\end{equation}
\[
D = \Big{(}\frac{1}{2}\cosh(\Omega\tau)+\frac{1}{2}\coth(r/2)\sinh(\Omega\tau)+ \frac{1}{2}
\]
\begin{equation}
+\frac{1}{2}\sinh(\Omega\tau)-\frac{1}{2}\coth(r/2)(\cosh(\Omega\tau)-1)\Big{)}^2+\sinh^2(\Omega\tau),
\end{equation}
and
\begin{equation}
E = (\bar{n}+1/2)\cosh(2r+2\Omega\tau)- \frac{1}{2}
\end{equation}
for $\theta =2 \varphi$.

The quantities $A$, $C$, $D$, and $E$ are all positive. However, $B$ may be either positive or negative. Accordingly, of the two possible solutions (B1), only the case with $B<0$ gives rise to $|\alpha|>0$, which corresponds to the solution (B1) with the upper sign. It is interesting that this occurs when $r>\frac{1}{2}\ln (2\bar{n}+1)$ for $\tau\geq 0$.

Solution (B1) simplifies considerably for the case $\tau=0$ in which case one has that the solution (B1) reduces to
\begin{equation}
|\alpha|=\frac{1}{2}\sqrt{\frac{(2\bar{n}+1)(e^{4r} -1) \big{(}(2\bar{n}+1)e^{2r} -1\big{)}}{e^{2r}-2\bar{n}-1}}.
\end{equation}
Note that such minimum occurs only for $r > \frac{1}{2}\ln(2\bar{n}+1)$. For $\bar{n}=0$ one has from (B7) that $|\alpha|=\frac{1}{2}\sqrt{(e^{4r} -1)}$ and the minimum correlation is given by
\begin{equation}
g^{(2)}_{(\bar{n}=0)}(0)= 2+ \frac{\sinh(2r)\big{(}\sinh(2r)-4|\alpha|^2 \big{ )}-4|\alpha|^4}{(\cosh(2r) -1 + 2|\alpha|^2)^2},
\end{equation}
where
\begin{equation}
r=\frac{1}{4}\ln(4|\alpha|^2+1).
\end{equation}
Results (B7)-(B9) agree with those of Ref. [4] found in their Appendix A. However, our result applies to the minimum value of $g^{(2)}(\tau)$ for $\tau>0$.

\end{document}